\documentclass{llncs}
\usepackage{amssymb}
\usepackage{bm}
\usepackage{subfigure}
\usepackage{amsmath}
\usepackage{amsbsy}
\usepackage{mathtools}
\usepackage{psfrag}
\usepackage{graphicx}
\usepackage{makeidx}
\usepackage{algorithm}
\usepackage{algorithmic}
\usepackage{subfigure}

\usepackage{verbatim}
\usepackage{amssymb}
\usepackage{amsfonts}
\usepackage{bm}

\ifx\@waldernographics\@empty
\relax
\else
\fi%

\newcommand{\psfragput}[4]{\psfrag{#1}{\begin{picture}(0,0)\put(#3,#4){#2}\end{picture}}}

\newcommand\tran{\top}

\newcommand{\ttran}{^{\tran}}

\newcommand{\chapternewpage}{~ \ifodd\value{page} \chapter*{} \fi}

\newcommand{\novelty}[3]{\ifthenelse{\equal{\value{#1}}{0}}{#3}{#2}\setcounter{#1}{1}}

\newcommand{\beqn}{\begin{equation}\/}
\newcommand{\eeqn}{\end{equation}\/}
\newcommand{\beqns}{\begin{equation*}\/}
\newcommand{\eeqns}{\end{equation*}\/}
\newcommand{\beqna}{\begin{eqnarray}\/}
\newcommand{\eeqna}{\end{eqnarray}\/}
\newcommand{\beqnas}{\begin{eqnarray*}\/}
\newcommand{\eeqnas}{\end{eqnarray*}\/}

\newcommand{\balign}{\begin{align}\/}
\newcommand{\ealign}{\end{align}\/}
\newcommand{\bals}{\begin{align*}\/}
\newcommand{\eals}{\end{align*}\/}

\newcommand{\Ocal}{\mathcal{O}}

\newcommand{\realset}{\mathbb{R}}

  {\begin{center}\begin{Sbox}\begin{minipage}}%
  {\end{minipage}\end{Sbox}\fbox{\TheSbox}\end{center}}

  {\begin{center}\begin{Sbox}\begin{minipage}}%
  {\end{minipage}\end{Sbox}\Ovalbox{\TheSbox}\end{center}}

\newcommand{\cmcolor}{}

\newcommand{\blocksperkernel}{\texttt{BlocksPerKernel}}
\newcommand{\threadsperblock}{\texttt{ThreadsPerBlock}}
\newcommand{\vperthread}{\texttt{ElementsPerThread}}

\begin{document}
\title{Rank $k$ Cholesky Up/Down-dating on the GPU:
\texttt{gpucholmodV0.2}}
\author{Christian Walder \\ ~ \\ November 4 2010}
\institute{Informatics and Mathematical Modelling \\
Technical University of Denmark, DK-2800 \\
\email{chwa@imm.dtu.dk}}  
\maketitle
\begin{abstract}
In this note we briefly describe our Cholesky modification algorithm for streaming multiprocessor architectures. Our implementation  is available in \textit{C++} with \textit{Matlab} binding, using \textit{CUDA} to utilise the graphics processing unit (GPU). Limited speed ups are possible due to the bandwidth bound nature of the problem. Furthermore, a complex dependency pattern must be obeyed, requiring multiple kernels to be launched. Nonetheless, this makes for an interesting problem, and our approach can reduce the computation time by a factor of around 7 for matrices of size $5000 \times 5000$ and $k=16$, in comparison with the LINPACK suite running on a CPU of comparable vintage. Much larger problems can be handled however due to the $\Ocal(n)$ scaling in required GPU memory of our method.
\end{abstract}
\section{Introduction and Problem Setting}
\label{SECproblem}
Given a symmetric positive definite matrix $A\in \realset^{n\times n}$, for reasons of computational efficiency and stability, it is often indispensable that we are able to maintain the upper triangular Cholesky factor $L$ such that $A=L\ttran L$ --- see \cite{seegerlowrank} for a discussion. Frequently, $A$ will changes by low rank modification during the course of an algorithm, hence it is imperative that we can accordingly modify the associated Cholesky factor $L$ in an efficient and stable manner, in order to maintain an optimal asymptotic time complexity. We focus on the following problem: given $A$, $L$, and a matrix $V \in \realset^{n\times k}$, form the modified factor $\tilde{L}$ such that $\tilde{L}\ttran\tilde{L} = A \pm V V\ttran = \tilde{A}$. In particular we do so with $\mathcal{O}(kn^2)$ operations, rather than by na{\"i}vely computing the modified matrix $\tilde{A}$ and from there rebuilding the full Cholesky factor. This is referred to as the rank $k$ Cholesky up (down) date when dealing with addition (subtraction) of $V V\ttran$. Existing CPU implementations such as \texttt{dchud} of LINPACK \cite{linpack} typically treat the case $k=1$. We allow $k > 1$ since this leads to more efficient memory access, although speedups are also obtainable with our algorithm for $k=1$ (naturally this demands a larger problem size $n$, however).
\section{Serial Algorithm}
\label{SECserial}
The serial algorithm which we will adapt to the GPU is the so called hyperbolic approach which we state as Algorithm \ref{ALGserial} for the case $k=1$. 
\begin{algorithm}[t]
 \caption{\texttt{CholeskymodifyX} \newline \label{ALGserial} Modify the Cholesky factor 
$L\in\realset^{n\times n}$ by $V\in \realset^n$, with $\sigma \in \pm 1$
being positive (negative) to specify an update (downdate).}
 \begin{algorithmic}[nothing]
  \STATE \texttt{\hspace{-4mm} // first alternative: }
  \STATE \texttt{\hspace{-4mm} function CholeskyModifyA}($L,V,\sigma$)
  \FOR{$i=1$ to $n$}
    \STATE \texttt{Compute}($c_i,s_i,L_{i,i},V_i,\sigma$)
    \FOR{$j=1$ to $i-1$}
      \STATE \texttt{Apply}($c_j,s_j,L_{j,i},V_i,\sigma$)
    \ENDFOR 
  \ENDFOR
 \end{algorithmic}
 \begin{algorithmic}[nothing]
  \STATE \texttt{\hspace{-4mm} // second alternative: }
  \STATE \texttt{\hspace{-4mm} function CholeskyModifyB}($L,V,\sigma$)
  \FOR{$i=1$ to $n$}
    \STATE \texttt{Compute}($c_i,s_i,L_{i,i},V_i,\sigma$)
    \FOR{$j=i+1$ to $n$}
      \STATE \texttt{Apply}($c_i,s_i,L_{i,j},V_j,\sigma$)
    \ENDFOR 
  \ENDFOR
 \end{algorithmic}
 \begin{algorithmic}[nothing]
  \STATE \texttt{\hspace{-4mm} // helper function: }
  \STATE \texttt{\hspace{-4mm} function Compute}($c_i,s_i,L_{i,i},V_i,\sigma$)
  \STATE $w \leftarrow \sqrt{L_{i,i}^2+\sigma V_i^2}$
  \STATE $c_i \leftarrow w / L_{i,i}$
  \STATE $s_i \leftarrow V_i / L_{i,i}$
  \STATE $L_{i,i} \leftarrow w$
 \end{algorithmic}
 \begin{algorithmic}[nothing]
  \STATE \texttt{\hspace{-4mm} // helper function: }
  \STATE \texttt{\hspace{-4mm} function Apply}($c_i,s_i,L_{i,j},V_j,\sigma$)
  \STATE $L_{i,j} \leftarrow (L_{i,j} + \sigma s_i V_j) / c_i$
  \STATE $V_j \leftarrow c_i V_j - s_i L_{i,j}$
 \end{algorithmic}
\end{algorithm}
\section{Parallel Version}
\label{SECparallel}
Let us take stock of the memory accesses in the inner loop of the two possible orderings in Algorithm \ref{ALGserial}:
\begin{itemize}
 \item In \texttt{CholeskyModifyA} we read $c_j,s_j$ and $L_{j,i}$, and write $L_{j,i}$. The $V_i$ must only be read and written before and after the inner loop.
 \item In \texttt{CholeskyModifyB} we read and write $L_{i,j}$ and $V_j$. In this case, it is $c_i$ and $s_i$ which need only be read and written before and after the inner loop.
\end{itemize}
Hence we see that if reading and writing were equally costly,  then the ordering of \texttt{CholeskyModifyA} would be slightly better --- not only that, but  \texttt{CholeskyModifyA} also offers a rather natural mapping to the GPU shared and register memory of current hardware, as we shall see in the following subsection, so this is the approach we will employ. We make no claim as to the optimality of this approach --- if the reader is aware of a superior approach, we would be interested to hear about it.
\section{Panelling}
\label{SECpanelling}
A GPU kernel function which computes the inner loop of \texttt{CholeskyModifyA} would need to be launched $n$ times. To avoid the overhead inherent in these repeated kernel launches we proceed by computing larger submatrices of $L$, sequentially and either on the CPU or the GPU. The panelling strategy, which is illustrated in Figure \ref{FIGpanels}, will be described in this section. 
\subsection{Parameters}
\begin{figure*}
\begin{center}
\includegraphics[width=0.45\textwidth]{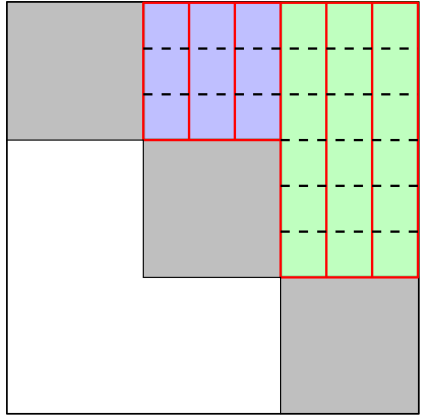}
\caption{\label{FIGpanels} A colour coding of the manner in which the panels of L are computed on either the CPU (grey panels) or the GPU (rest) --- see section \ref{SECpanelling} for an explanation.}
\end{center}
\end{figure*}

The algorithm has the following parameters:
\begin{enumerate}
 \item \blocksperkernel, which is 3 in figure \ref{FIGpanels} and 28 in our implementation.
 \item \threadsperblock, which is 32 in our implementation and unspecified in the figure (but must equal $n / (3 \times \text{\blocksperkernel})$ as the figure depicts three diagonal chunks).
 \item \vperthread, which is 16 in our implementation and unspecified in the figure. This parameter corresponds to the number of columns of $V$ which we process in each kernel call. That is, $k / \text{\vperthread}$ successive kernel calls will be employed to process the entire update matrix $V$ in batches of size \vperthread.
\end{enumerate}
We now describe the r{\^o}les of the CPU and GPU in their respective phases of the computation. We do not provide a detailed description but rather a high level overview which could serve as an aid in deciphering our C++ implementation.
\subsection{Panel Ordering}
The panels are processed in the order top-left grey: CPU; blue: GPU; middle gray: CPU; green: GPU; and bottom-right grey: CPU.
\subsection{CPU --- On Diagonal Sub-matrices}
\label{SECcpublock}
The grey blocks in the figure are of size \blocksperkernel\ $\times$ \threadsperblock, and are calculated on the CPU. This is trivially computed via  Algorithm \ref{ALGserial} combined with a loop over the \vperthread\ update vectors.
\subsection{GPU --- Off Diagonal Sub-matrices}
\label{SECgpublock}
\subsubsection{Upload to the GPU}
Before launching the kernel we transmit the elements of the $c$ and $s$ vectors from the previous CPU submatrix, as well as the sub-matrix of $L$ corresponding to the current GPU iteration. 
\subsubsection{Compute on the GPU}
The GPU kernel is divided into \blocksperkernel\ blocks (the threads in each of which may communicate via shared memory, as is the nature of the streaming multiprocessor architecture). Each GPU block computes a rectangular sub-matrix of $L$ --- these are marked by the tall, skinny red rectangles in the figure. Abstractly speaking, each block proceeds as follows:
\begin{enumerate}
 \item Load the elements of $V$ corresponding to the columns of the current red rectangle into local per-thread registers. For each column of $L$ we load \vperthread\ elements of $V$ --- each thread handles one column of $L$.
 \item Iterate over the square \threadsperblock\ $\times$ \threadsperblock\ sized sub-matrices delineated by dotted lines:
\begin{enumerate}
 \item Load the elements of $c$ and $s$ corresponding to the rows of the current square sub-matrix into per-block shared memory. Since we operate on batches of $V$, there will be \vperthread\ per row.
 \item Iterate over the rows of the square sub-matrix:
\begin{enumerate}
 \item read an element of $L$;
 \item call the \texttt{Apply} function (\vperthread\ times); 
 \item write that element of $L$ back into global memory.
\end{enumerate}
\end{enumerate}
\item Write the elements of $V$ from the first step back to global GPU memory.
\end{enumerate}
Note that this algorithm requires twice as much shared memory as registers. Happily that is also the ratio of the amount of those memories available on the latest devices.
\subsubsection{Download from the GPU}
On completion of the kernel, we send the appropriate submatrices of $L$ and $V$ back to host memory. 
\section{Results}
To give the reader an idea of roughly what our GPU algorithm might buy them in terms of speedups, we present some basic timings in figures \ref{FIGsixteen} and \ref{FIGone}. The experimental procedure involves forming the matrices $B\in\realset^{n\times n}$ and $V \in \realset^{n\times k}$ with elements drawn i.i.d. from the uniform distribution on $[0,1]$. In the update test we let $A = B\ttran B + I$, where $I$ is the identity matrix, and compute the Cholesky factor $L$ using the LAPACK algorithm \cite{lapack} and update it by $V$. For the downdating test we let $A = B\ttran B + I + V V\ttran$ and downdate by $V$. Errors are calculated as $\text{max}_{i,j} \left| \tilde{A}_{i,j} - C_{i,j}\right|$ where $C = \tilde{L}\ttran \tilde{L}$ is computed with the BLAS and $\tilde{L}$ is computed from $L$ by up/down-dating $L$. The experiment is repeated for $k=16$ and $k=1$ in figures \ref{FIGsixteen} and \ref{FIGone}, respectively. For the CPU up/down-dating we used the LAPACK suite. 

The test system was a desktop machine running 64 bit Ubuntu linux with a 2.8GHz Intel i7 CPU and an Nvidia Tesla C2050 GPU with 14 streaming multiprocessors and a total of 448 cores and CUDA version 3.10. Note that the number of cores on the GPU is relatively unimportant however, due to the bandwidth bound nature of the problem.

The experiments show that for $k=16$, the GPU overtakes the CPU at around $n=2000$, while for $k=1$ we require at least $n=4000$ in order to break even with the CPU. The errors are always very similar. Note that much larger problems can be handled since we only store $\Ocal(n)$ sized panels of $L$ in device memory, unfortunately however our test system had rather limited host memory.
\newcommand{\figwidth}{4.8cm}
\newcommand{\vertoff}{1}
\begin{figure}
\begin{center}
  \subfigure[single precision update]{
    \psfragput{xlabel}{$n$}{0}{-3}
    \psfragput{ylabel}{time (s)}{-10}{\vertoff}
    \includegraphics[width=\figwidth]{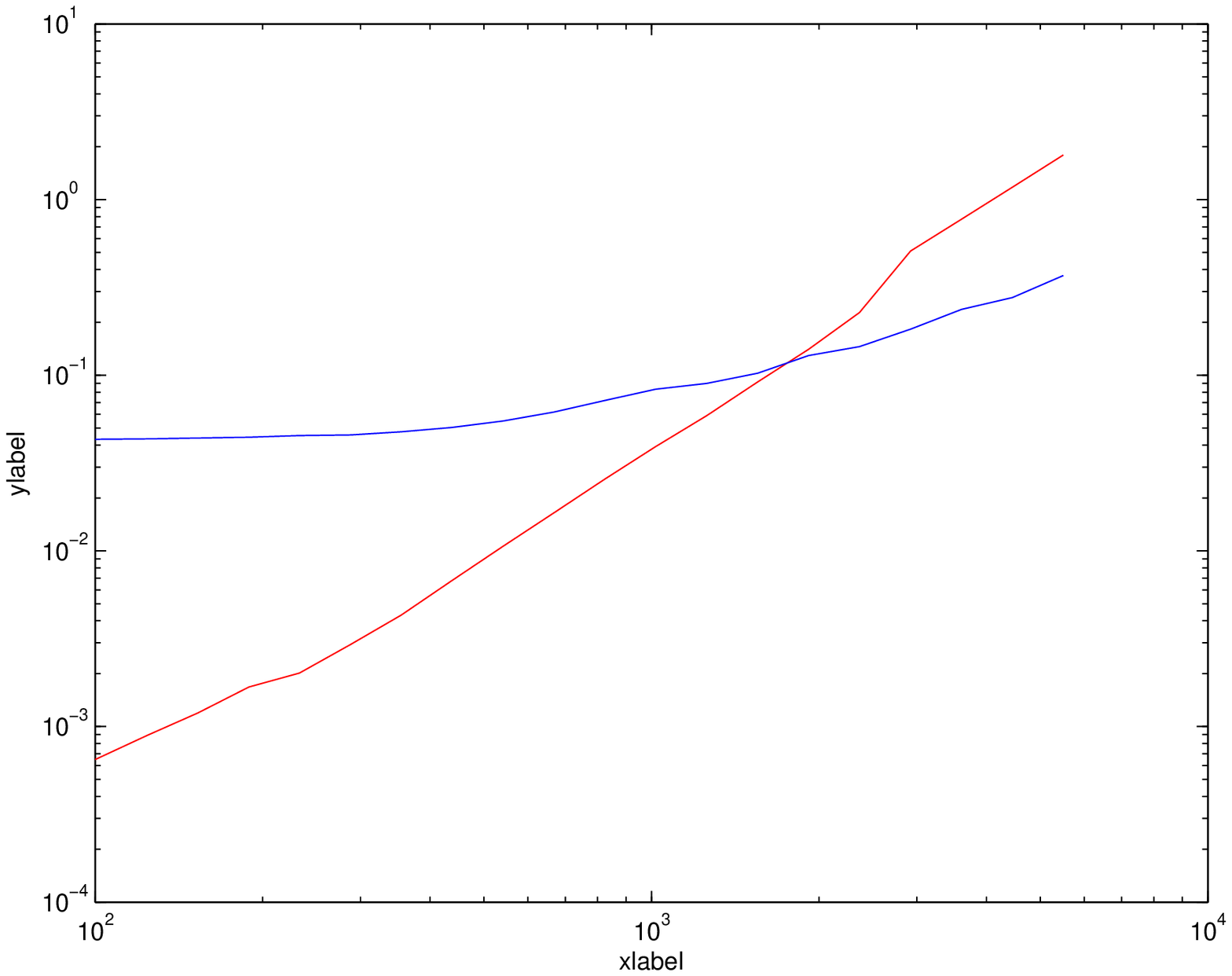}
  }
  \subfigure[single precision update]{
    \psfragput{xlabel}{$n$}{0}{-3}
    \psfragput{ylabel}{error}{-10}{\vertoff}
    \includegraphics[width=\figwidth]{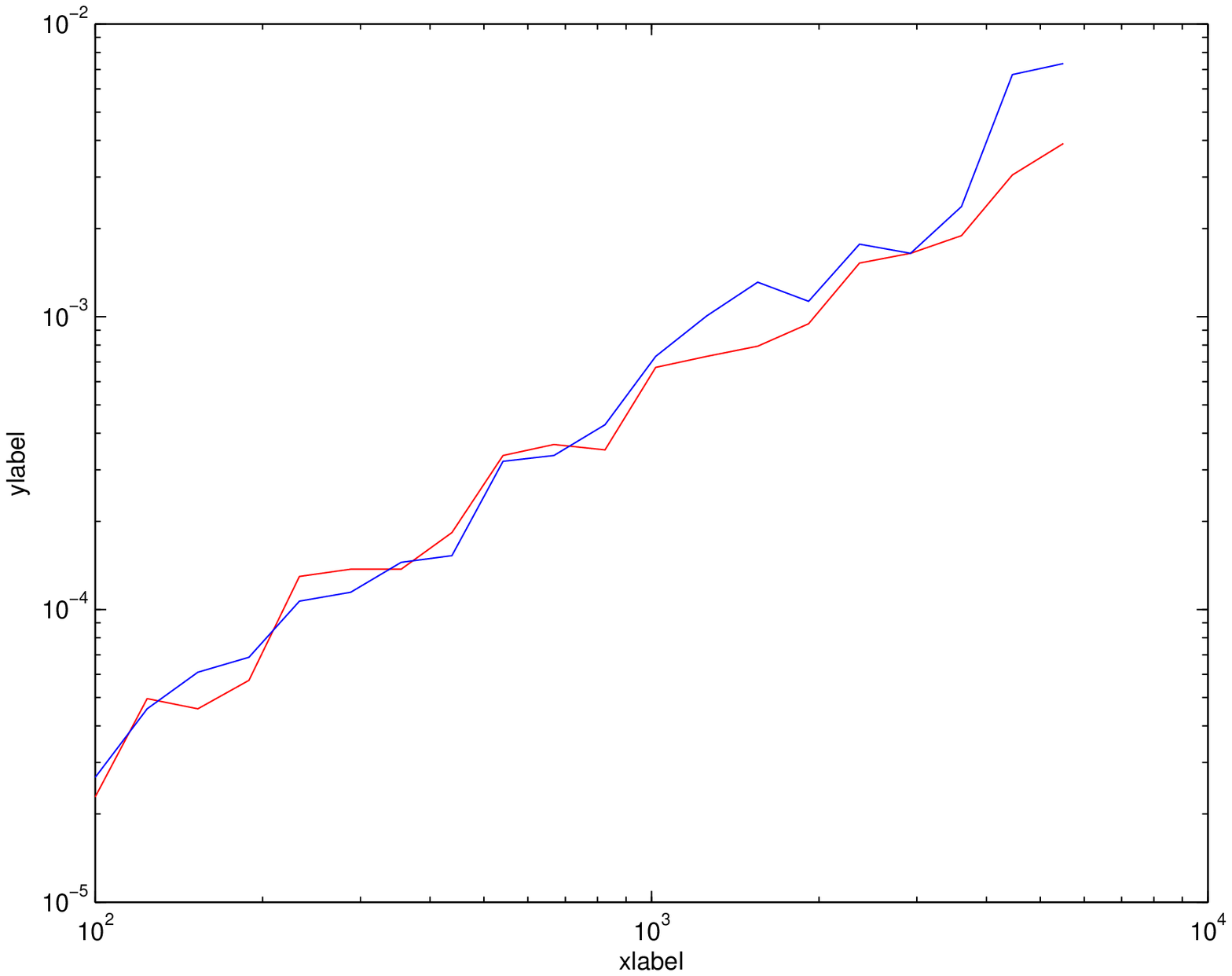}
  }
  \subfigure[single precision downdate]{
    \psfragput{xlabel}{$n$}{0}{-3}
    \psfragput{ylabel}{time (s)}{-10}{\vertoff}
    \includegraphics[width=\figwidth]{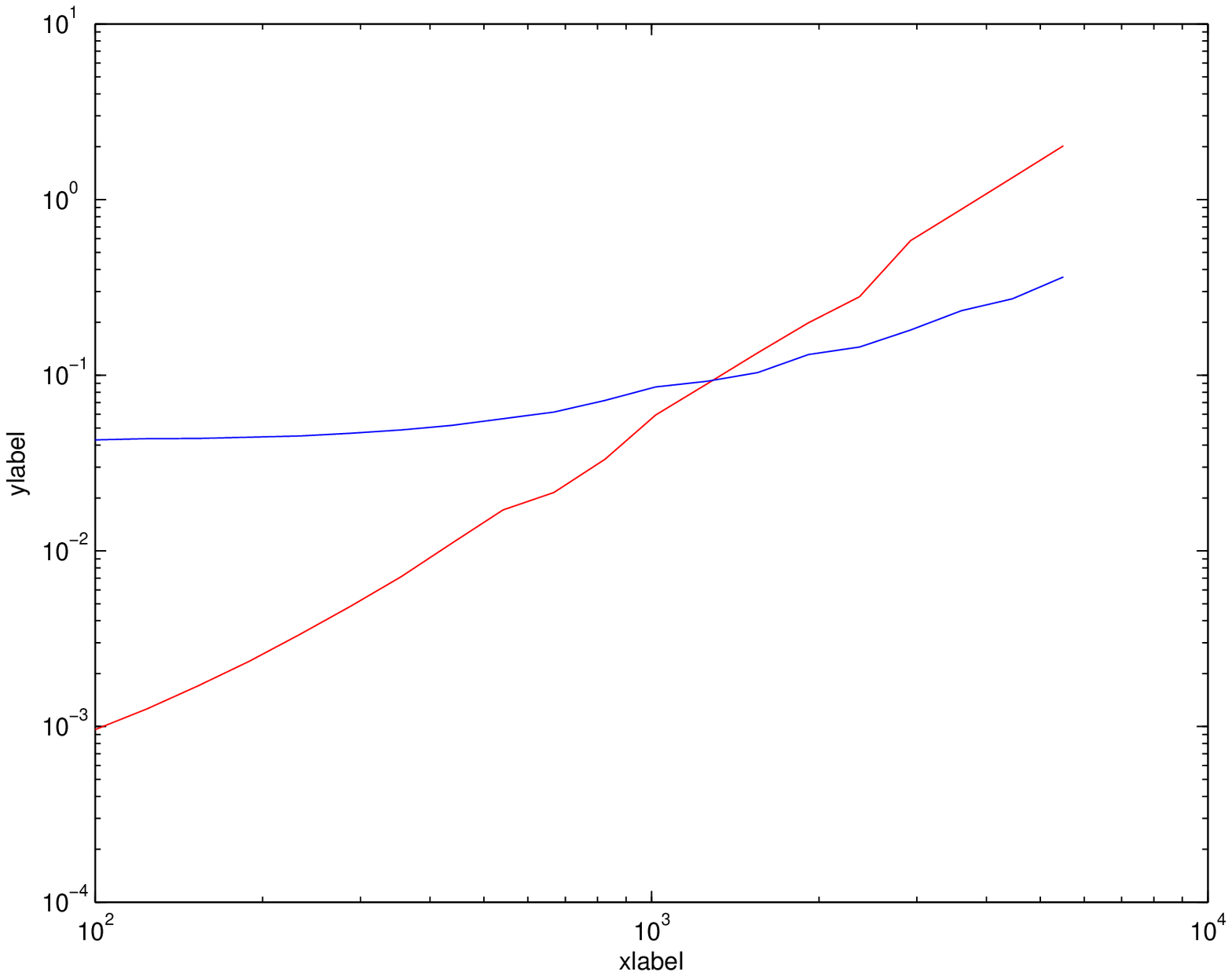}
  }
  \subfigure[single precision downdate]{
    \psfragput{xlabel}{$n$}{0}{-3}
    \psfragput{ylabel}{error}{-10}{\vertoff}
    \includegraphics[width=\figwidth]{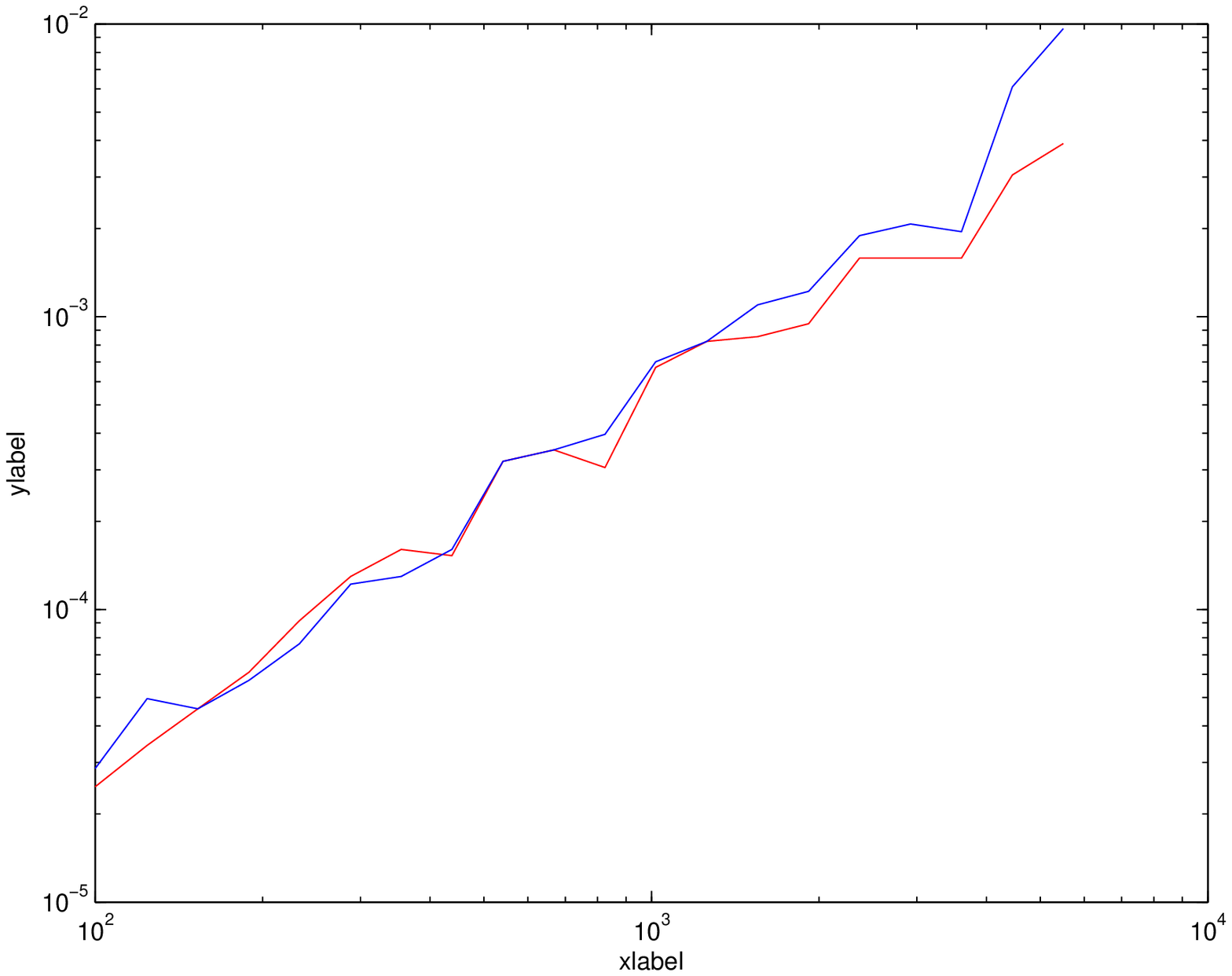}
  }
  \subfigure[double precision update]{
    \psfragput{xlabel}{$n$}{0}{-3}
    \psfragput{ylabel}{time (s)}{-10}{\vertoff}
    \includegraphics[width=\figwidth]{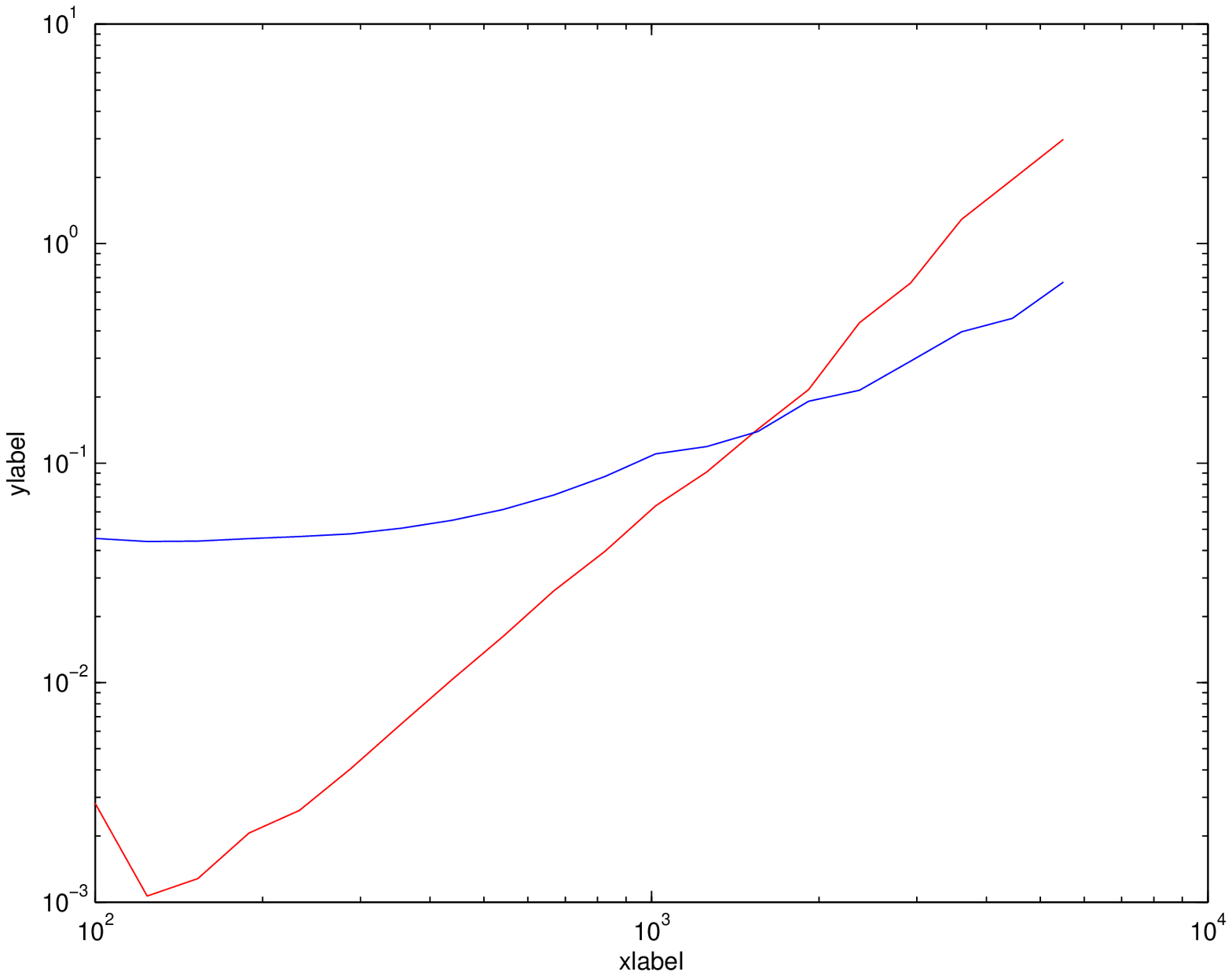}
  }
  \subfigure[double precision update]{
    \psfragput{xlabel}{$n$}{0}{-3}
    \psfragput{ylabel}{error}{-10}{\vertoff}
    \includegraphics[width=\figwidth]{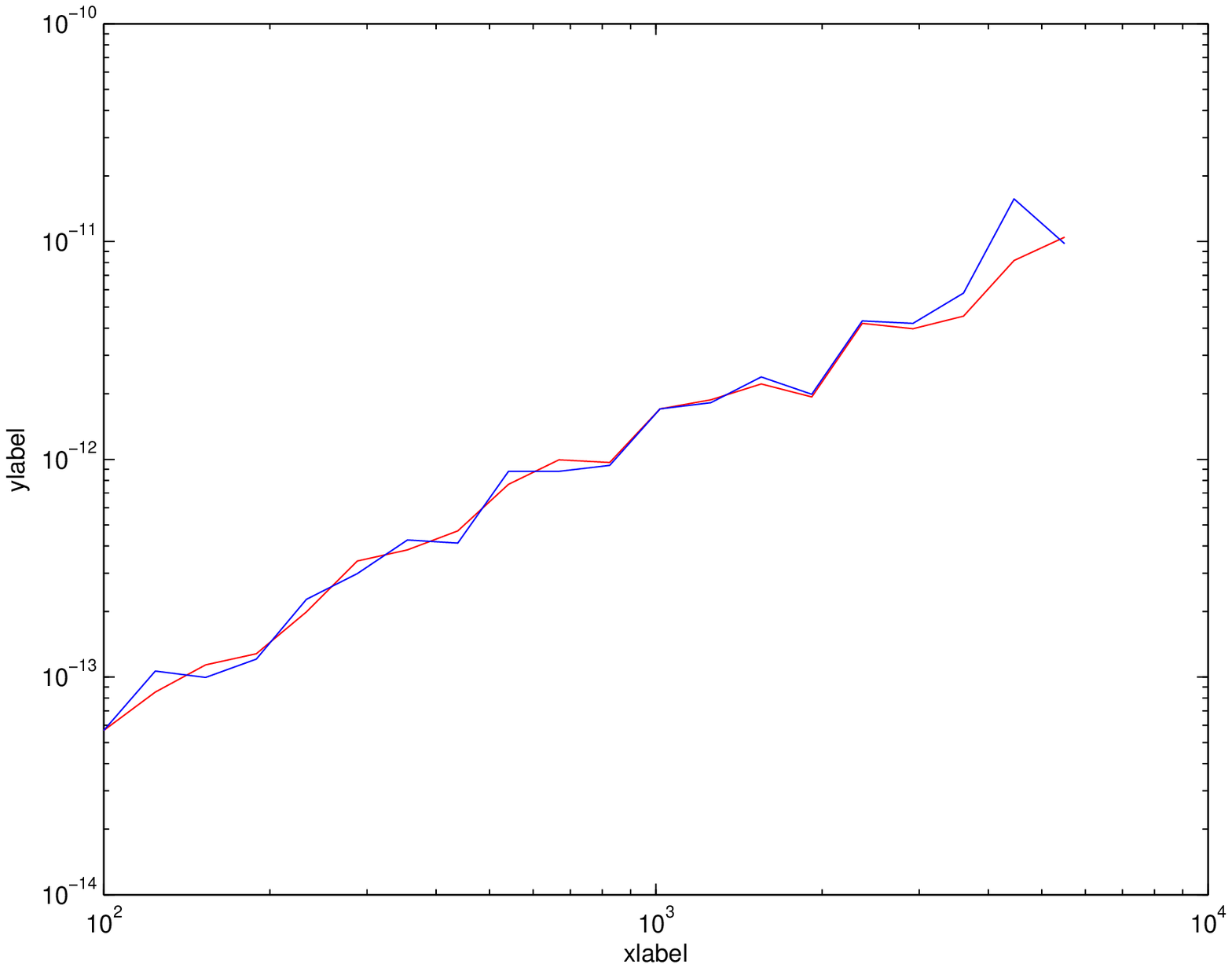}
  }
  \subfigure[double precision downdate]{
    \psfragput{xlabel}{$n$}{0}{-3}
    \psfragput{ylabel}{time (s)}{-10}{\vertoff}
    \includegraphics[width=\figwidth]{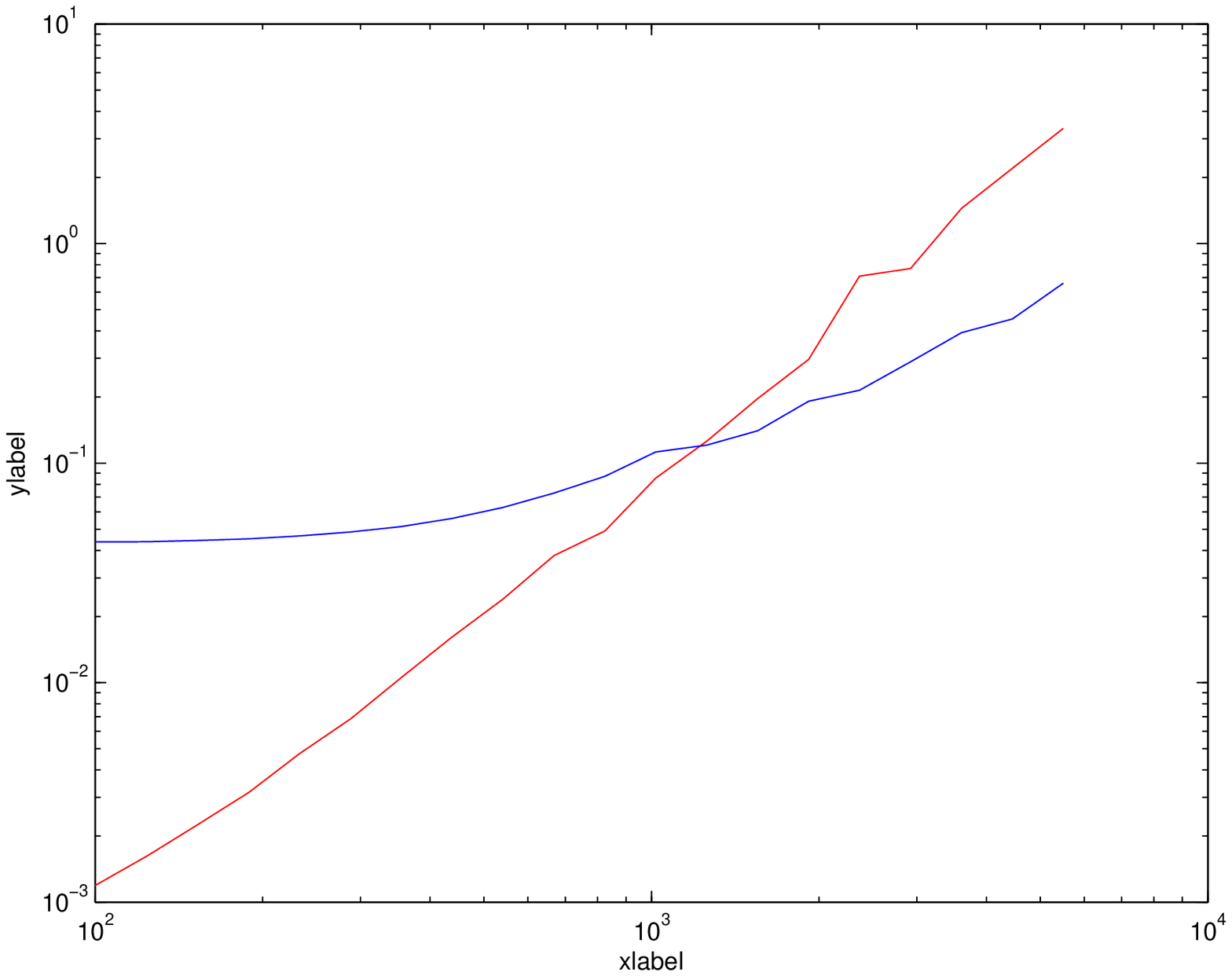}
  }
  \subfigure[double precision downdate]{
    \psfragput{xlabel}{$n$}{0}{-3}
    \psfragput{ylabel}{error}{-10}{\vertoff}
    \includegraphics[width=\figwidth]{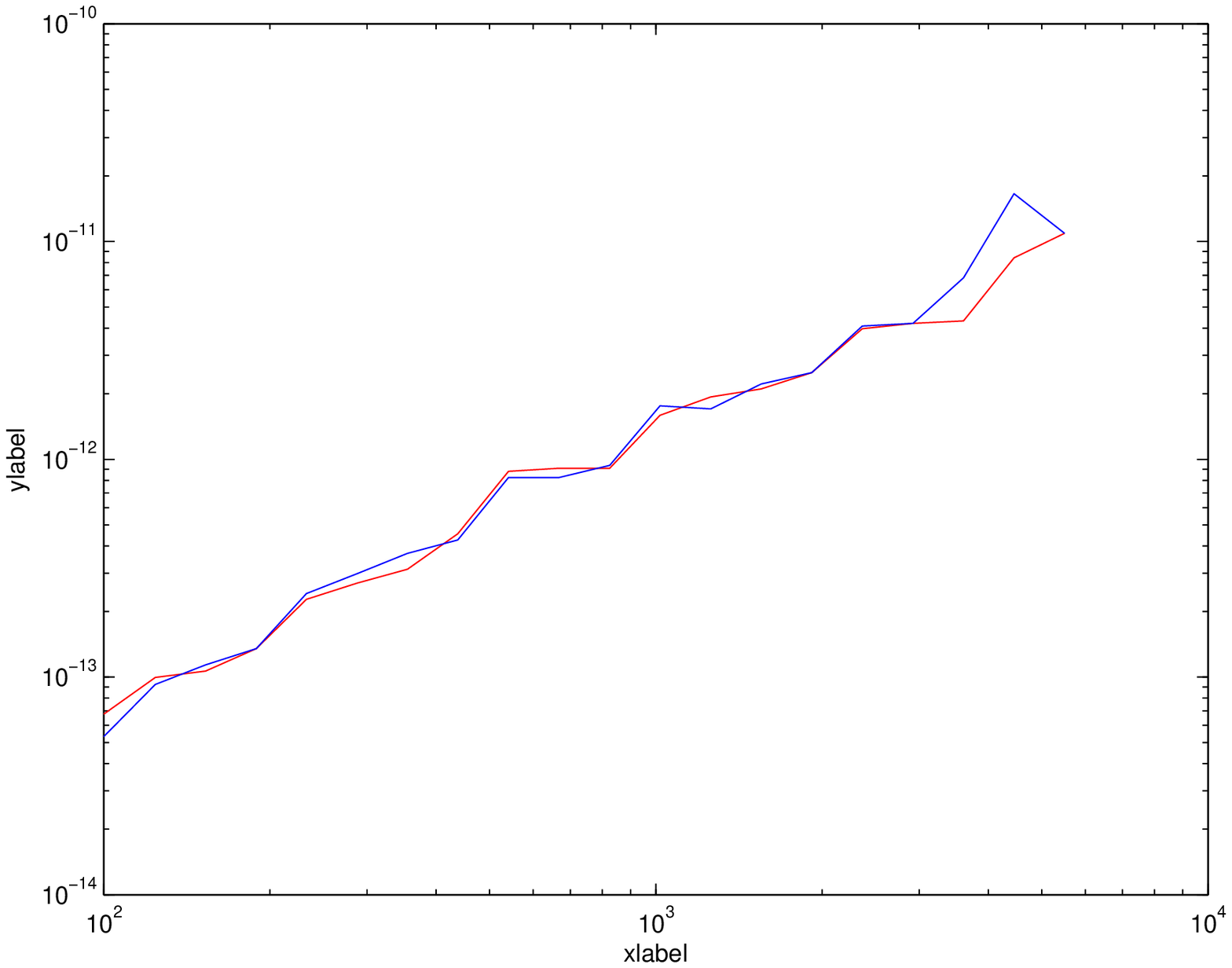}
  }
\end{center}
\caption{\label{FIGsixteen} Timings and errors for $k=16$ on the CPU (red) and GPU (blue).}
\end{figure}
\begin{figure}
\begin{center}
  \subfigure[single precision update]{
    \psfragput{xlabel}{$n$}{0}{-3}
    \psfragput{ylabel}{time (s)}{-10}{\vertoff}
    \includegraphics[width=\figwidth]{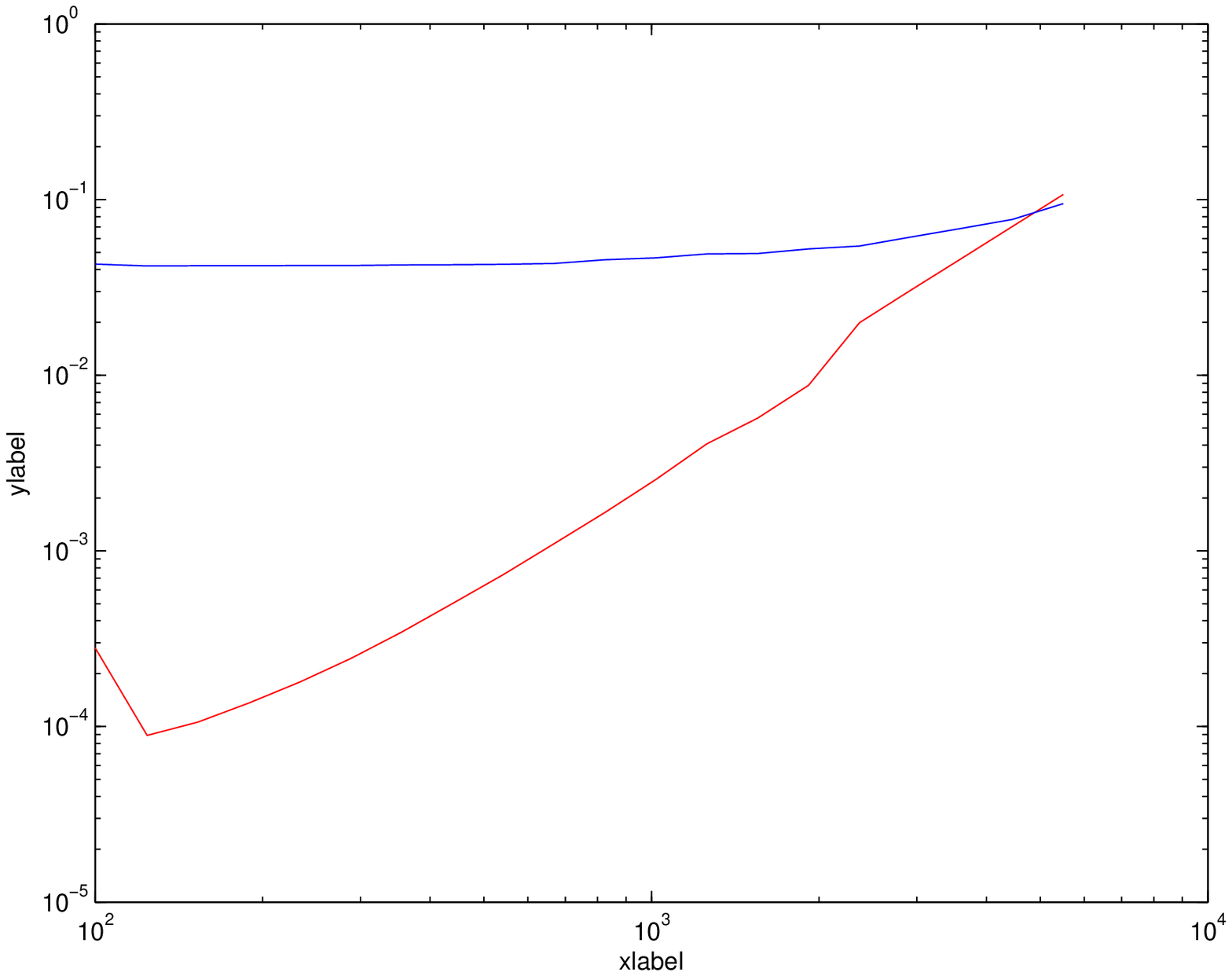}
  }
  \subfigure[single precision update]{
    \psfragput{xlabel}{$n$}{0}{-3}
    \psfragput{ylabel}{error}{-10}{\vertoff}
    \includegraphics[width=\figwidth]{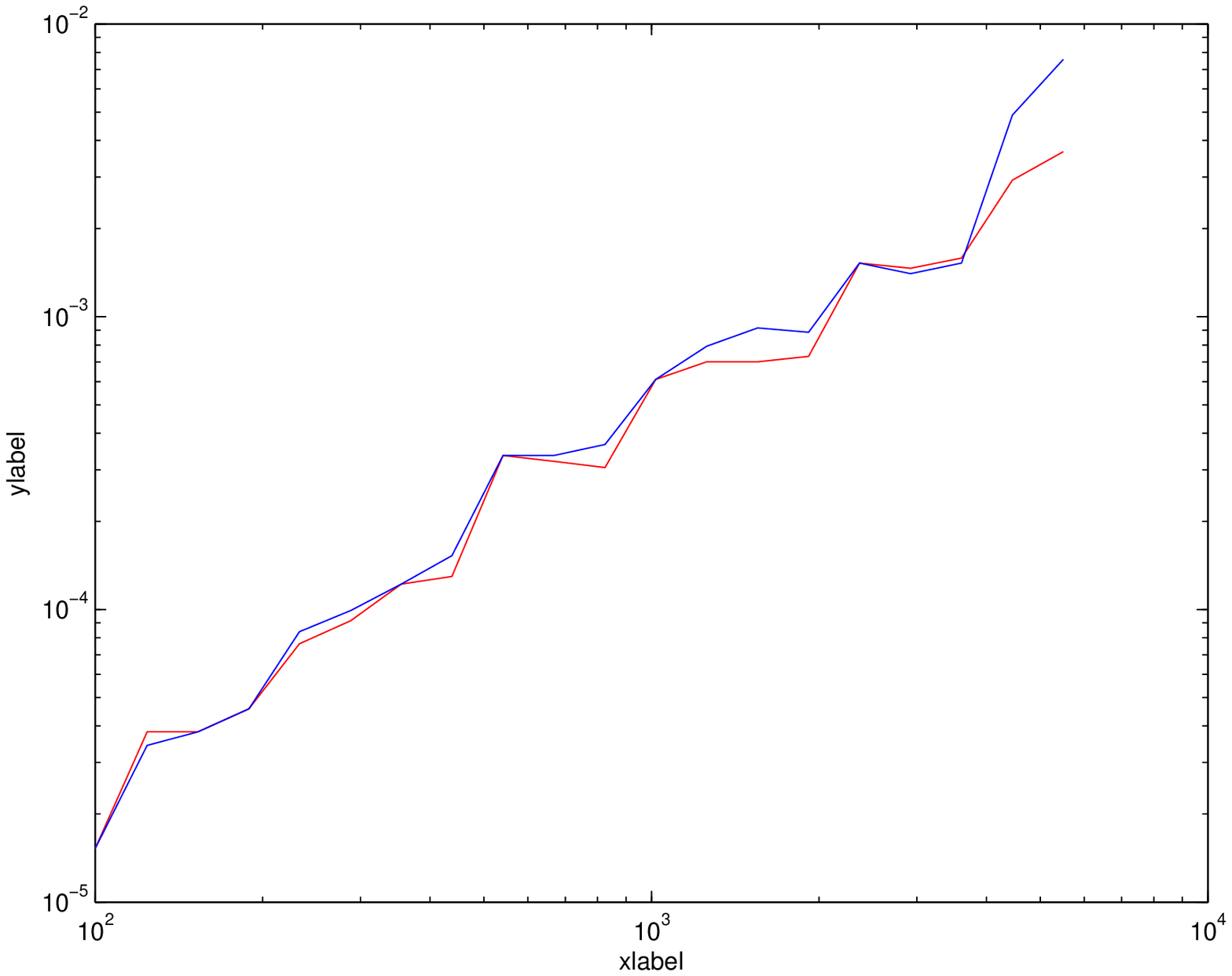}
  }
  \subfigure[single precision downdate]{
    \psfragput{xlabel}{$n$}{0}{-3}
    \psfragput{ylabel}{time (s)}{-10}{\vertoff}
    \includegraphics[width=\figwidth]{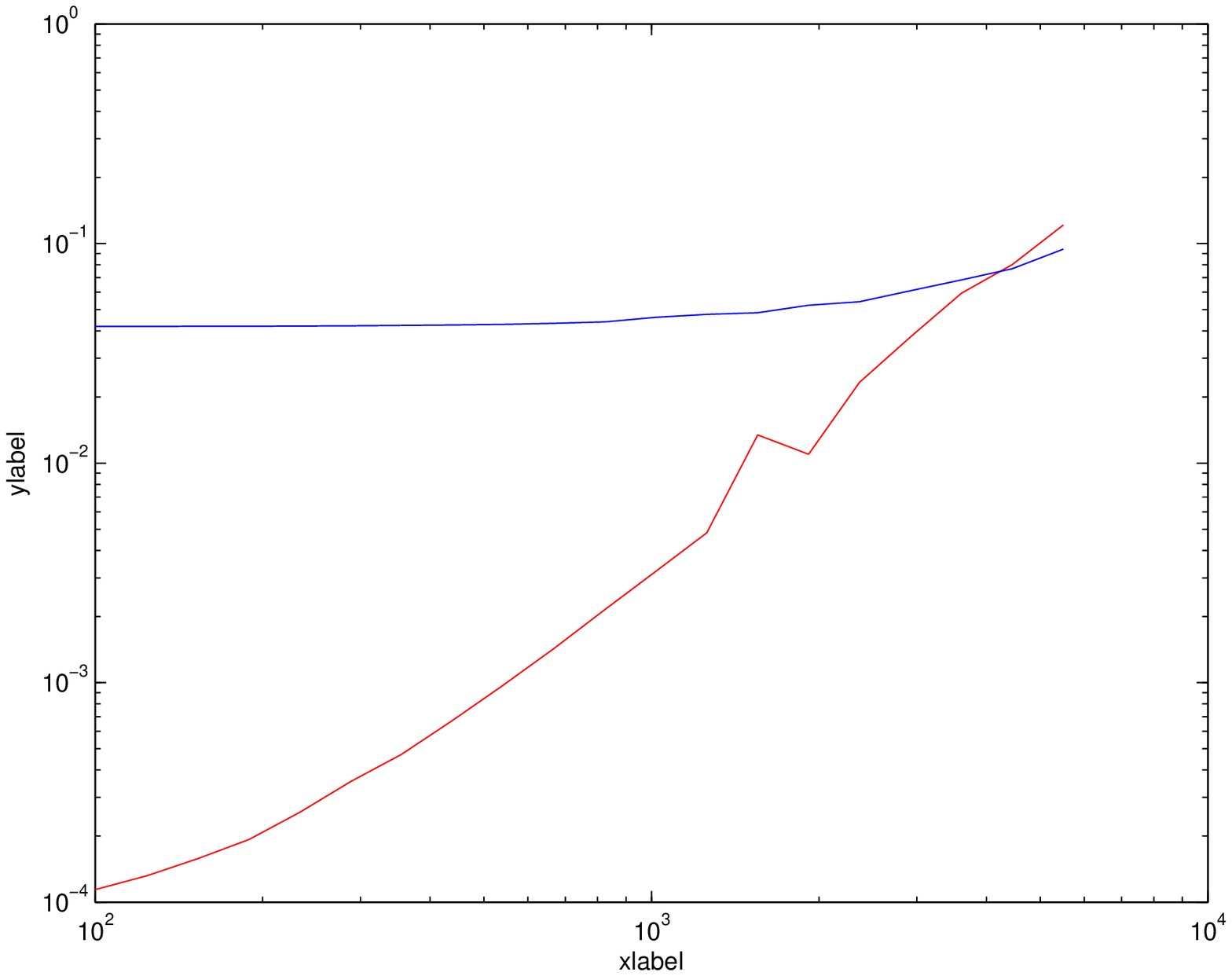}
  }
  \subfigure[single precision downdate]{
    \psfragput{xlabel}{$n$}{0}{-3}
    \psfragput{ylabel}{error}{-10}{\vertoff}
    \includegraphics[width=\figwidth]{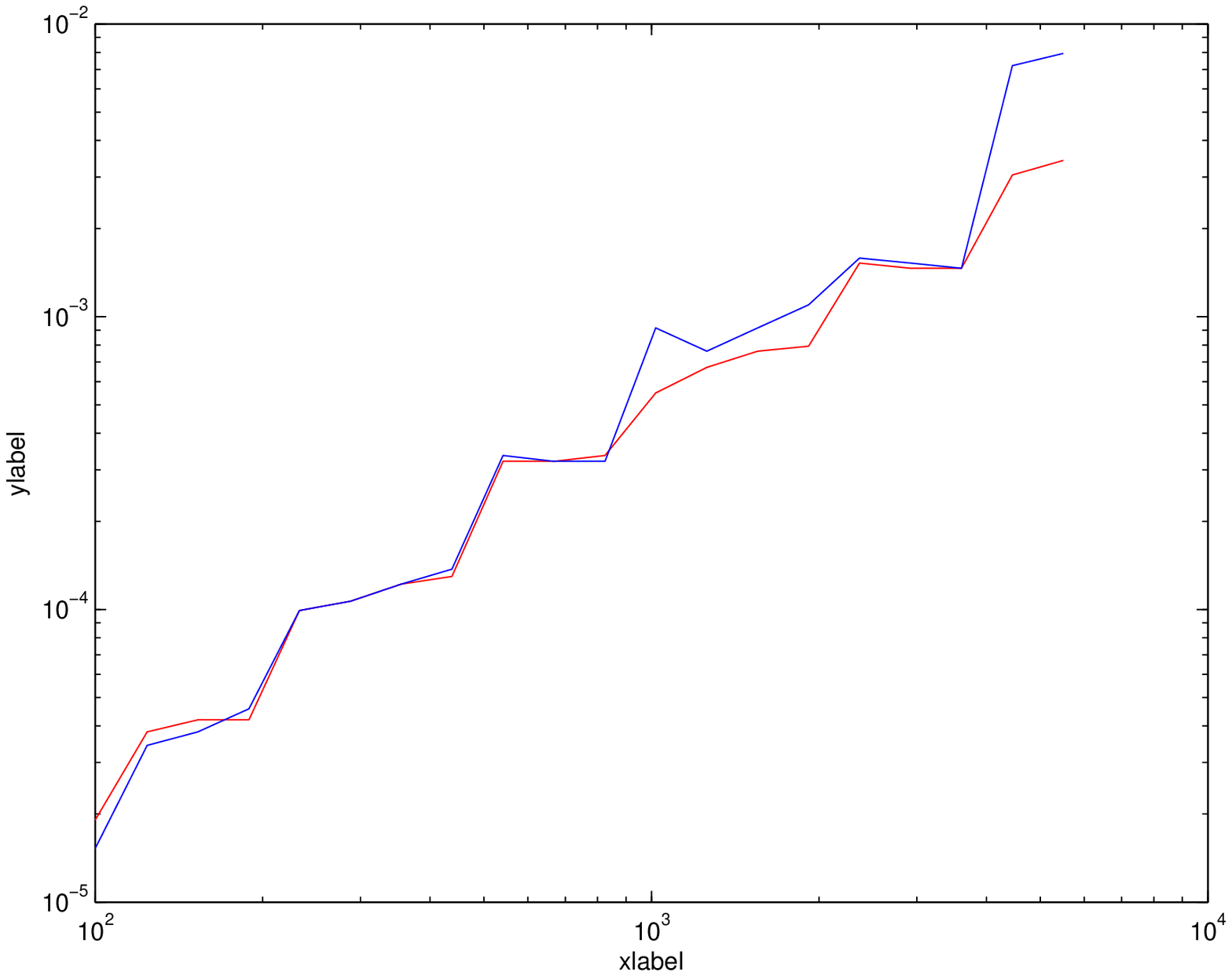}
  }
  \subfigure[double precision update]{
    \psfragput{xlabel}{$n$}{0}{-3}
    \psfragput{ylabel}{time (s)}{-10}{\vertoff}
    \includegraphics[width=\figwidth]{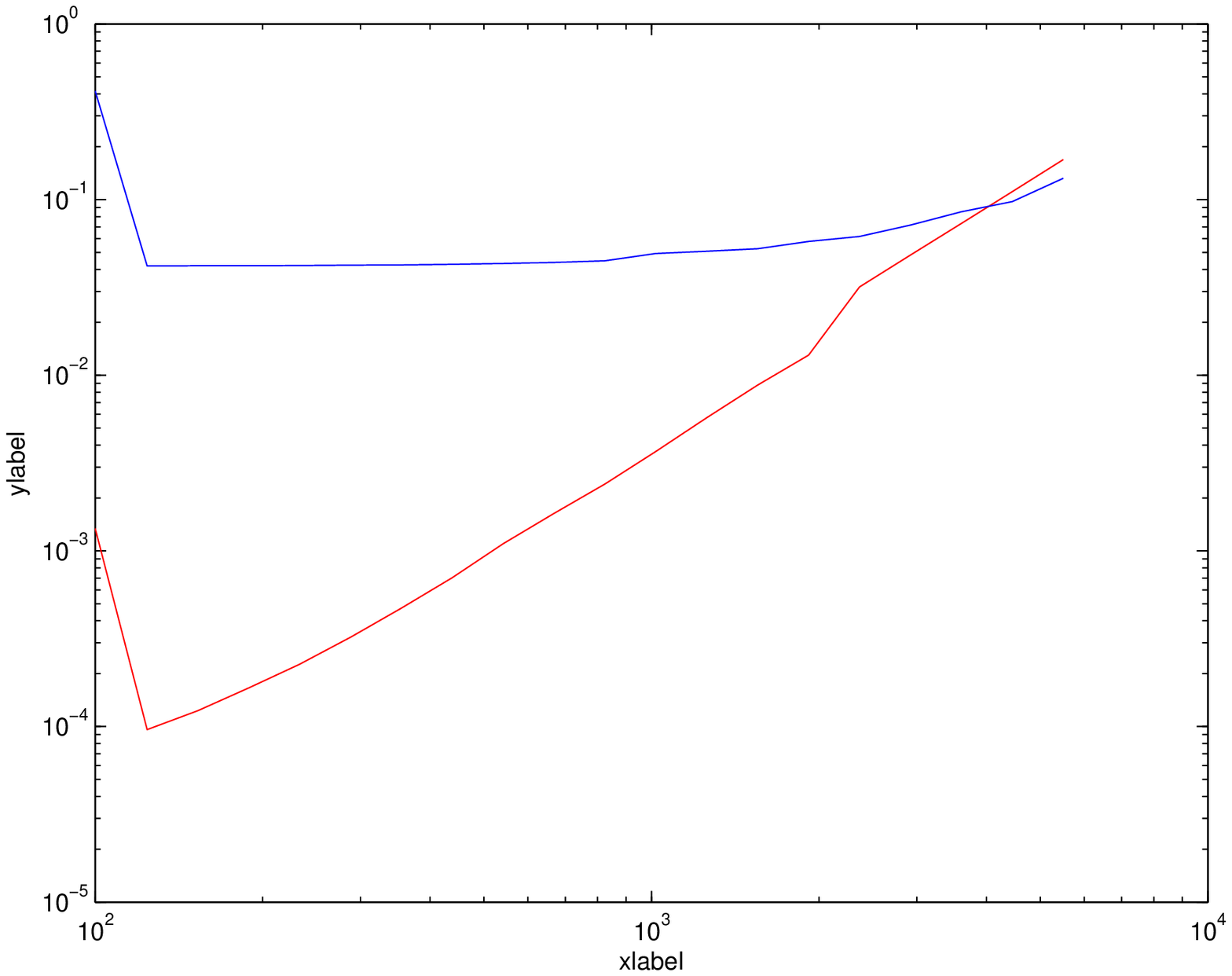}
  }
  \subfigure[double precision update]{
    \psfragput{xlabel}{$n$}{0}{-3}
    \psfragput{ylabel}{error}{-10}{\vertoff}
    \includegraphics[width=\figwidth]{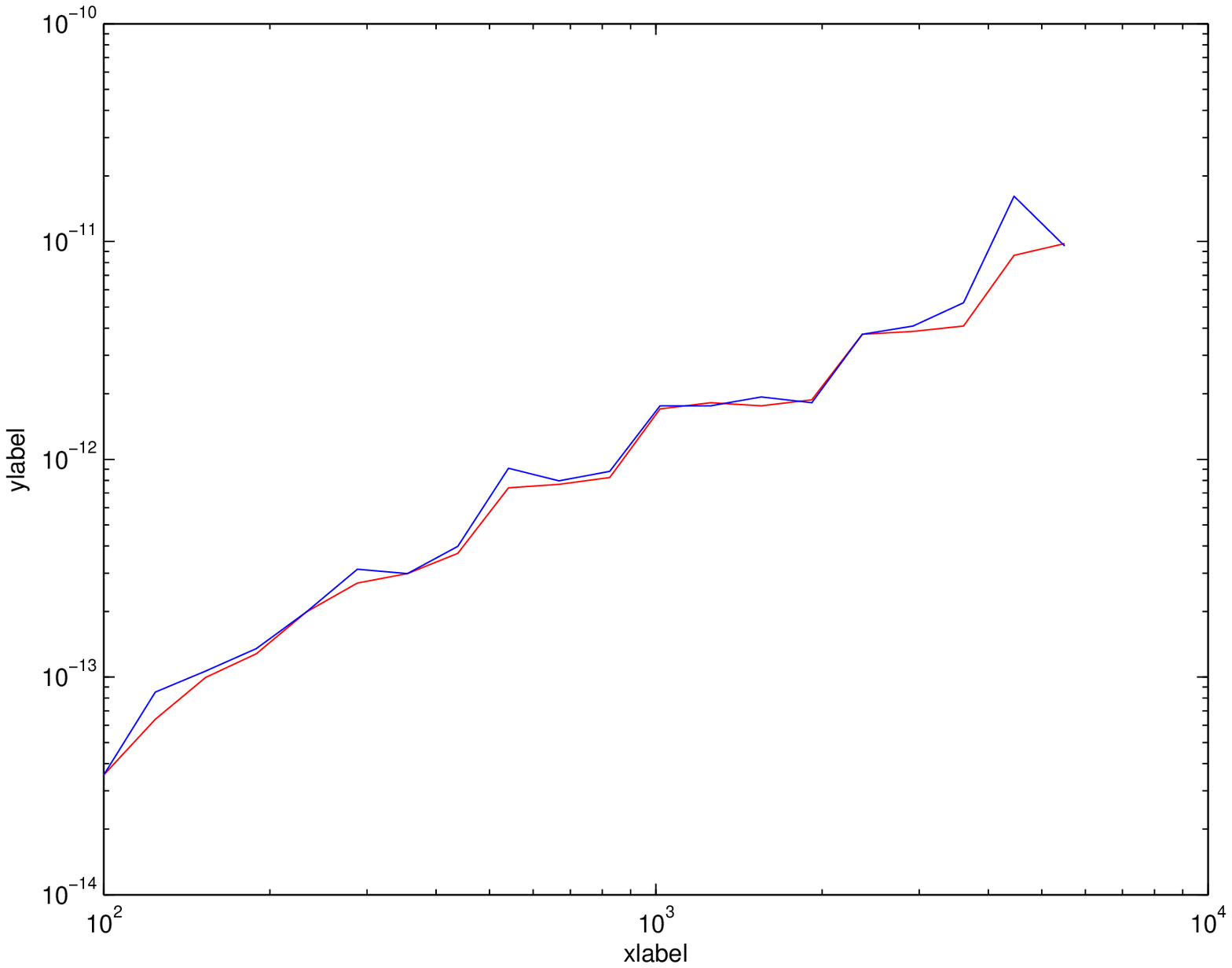}
  }
  \subfigure[double precision downdate]{
    \psfragput{xlabel}{$n$}{0}{-3}
    \psfragput{ylabel}{time (s)}{-10}{\vertoff}
    \includegraphics[width=\figwidth]{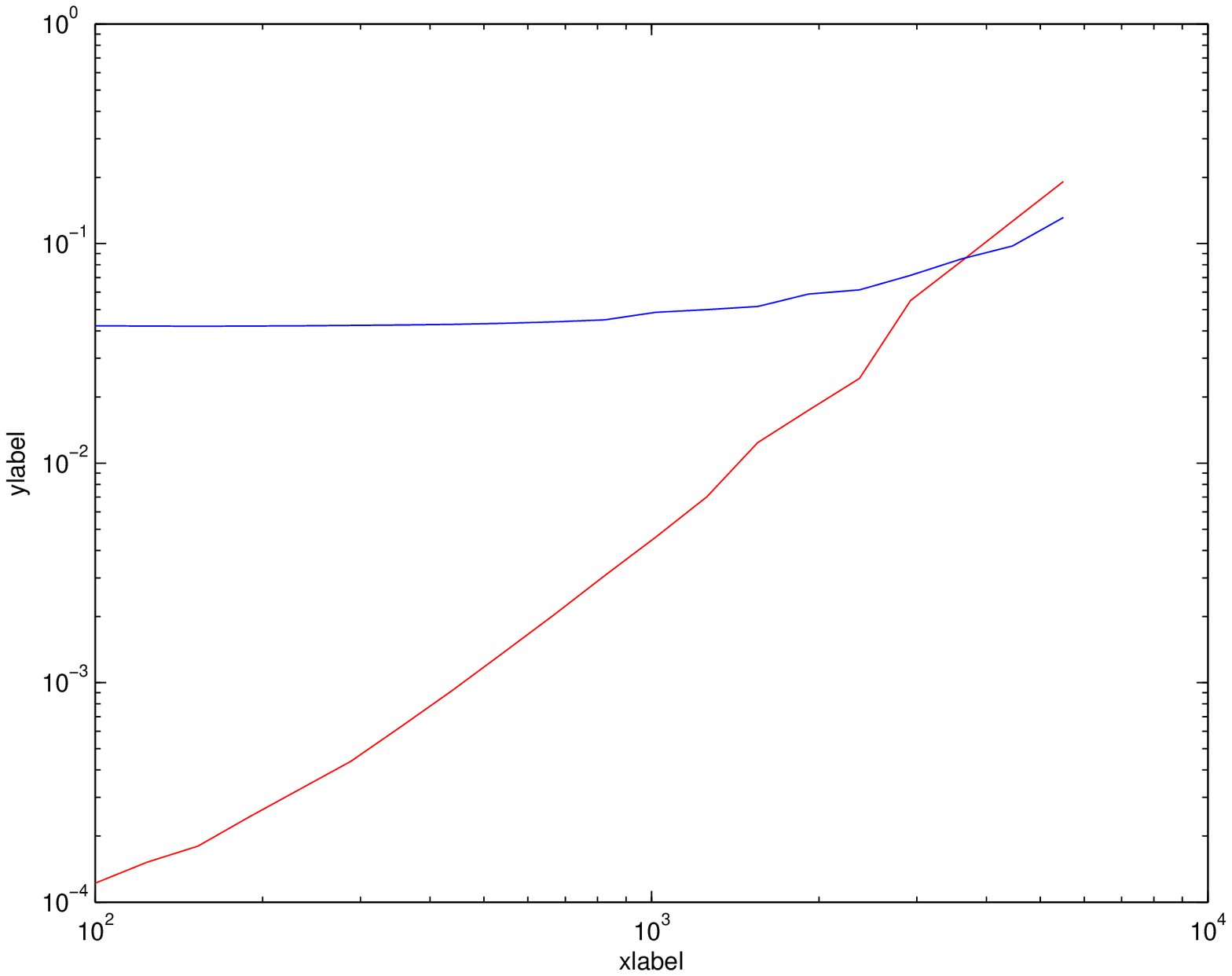}
  }
  \subfigure[double precision downdate]{
    \psfragput{xlabel}{$n$}{0}{-3}
    \psfragput{ylabel}{error}{-10}{\vertoff}
    \includegraphics[width=\figwidth]{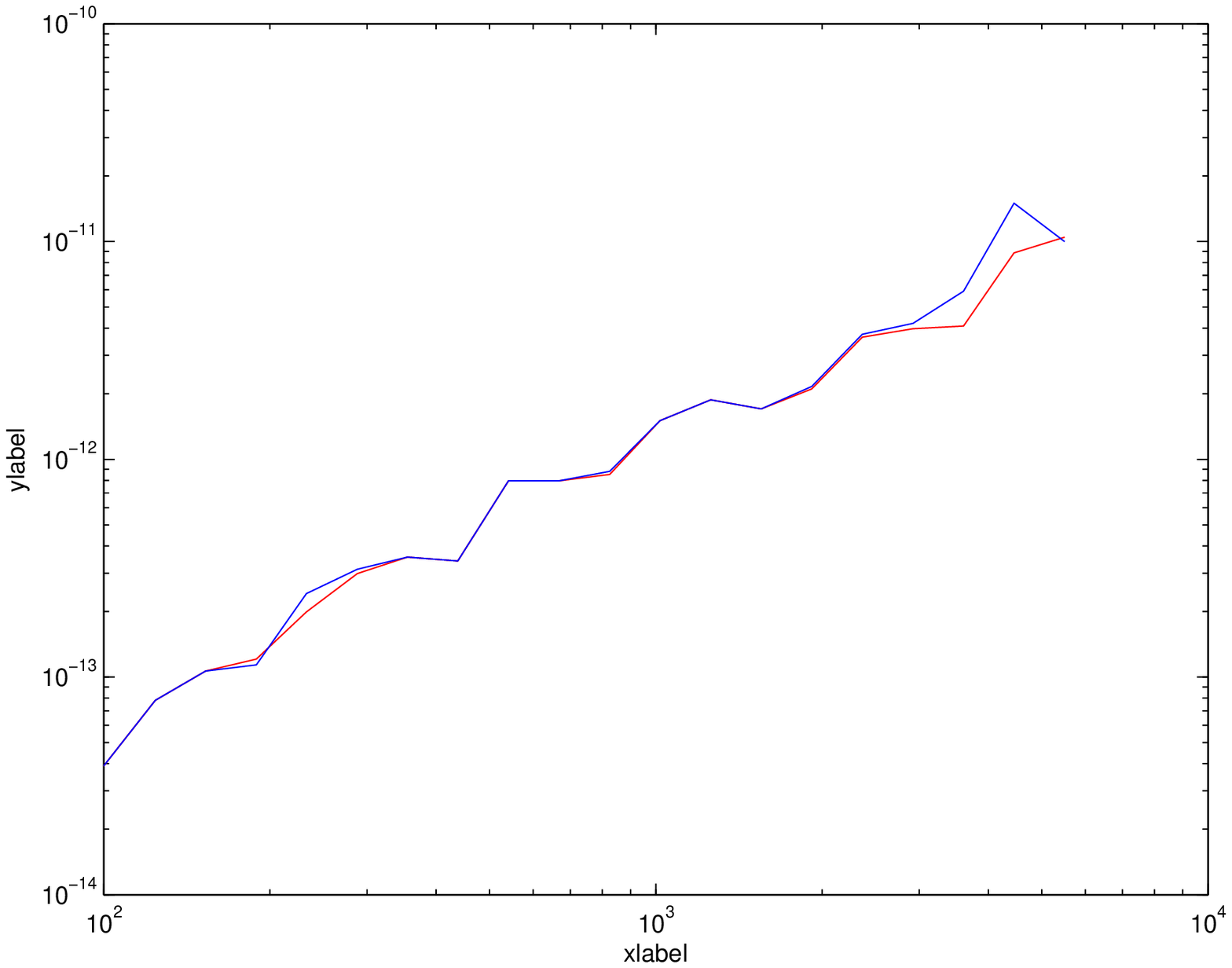}
  }
\end{center}
\caption{\label{FIGone} Timings and errors for $k=1$ on the CPU (red) and GPU (blue).}
\end{figure}
\bibliographystyle{splncs}
\bibliography{walder}
\end{document}